\documentclass[twocolumn,aps,amssymb,prl,showpacs]{revtex4}
\usepackage{graphicx}
\usepackage{amsmath}

\newcommand{\m}{\mathrm}

 \bigskip
\begin{document}
\title{Autler-Townes effect in a superconducting three-level system}

\author{Mika A. Sillanp\"a\"a$^1$}
\email{Mika.Sillanpaa@iki.fi}
\author{Jian Li$^1$}
\author{Katarina Cicak$^2$}
\author{Fabio Altomare$^2$}
\author{Jae I. Park$^2$}
\author{Raymond W. Simmonds$^2$}
\author{G. S. Paraoanu$^1$}
\author{Pertti J. Hakonen$^1$}

\affiliation{$^1$Helsinki University of Technology, Low Temperature
Laboratory, Puumiehenkuja 2B, Espoo, FIN-02015 HUT Finland \\
$^2$National Institute of Standards and Technology, 325
Broadway, Boulder CO 80305, USA}

\begin{abstract}
When a three-level quantum system is irradiated by an intense coupling field resonant with one of the three possible transitions, the absorption peak of an additional probe field involving the remaining level is split. This process is known in quantum optics as the Autler-Townes effect. We observe these phenomena in a superconducting Josephson phase qubit, which can be considered an "artificial atom" with a multilevel quantum structure. The spectroscopy peaks can be explained reasonably well by a simple three-level Hamiltonian model. Simulation of a more complete model (including dissipation, higher levels, and cross-coupling) provides excellent agreement with all the experimental data.
\end{abstract}

\maketitle

Superconducting qubits \cite{nakamuraqb} have recently been employed as a testing ground for quantum mechanics in a nearly macroscopic system. The interaction of the effective two-level system with a resonant cavity has attracted a lot of attention \cite{YaleQED,bus,MartinisFock,VacSplit}. However, Josephson junction-based quantum systems also possess higher quantum states or energy levels, beyond the two-level approximation, and these have received much less attention. For driven Rabi oscillations between the lowest two levels of a phase qubit, these higher states can reduce the apparent Rabi frequency at large drive powers \cite{Buisson,Wellstood} as the energy leaks out of the subspace spanned by the ground state $|0\rangle$ and the first excited state $|1\rangle$. Recently, the second excited state $|2\rangle$ has been used to characterize quantum bit (qubit) gate operations \cite{MartinisGate08}.

In atomic physics and quantum optics, various experiments have utilized the naturally occurring multilevel state structure of "real" atoms. A system accessing just three energy levels can display a rich variety of phenomena. Coherent population trapping, electromagnetically induced transparency (EIT) \cite{EIT91}, Autler-Townes splitting \cite{AT}, and stimulated Raman adiabatic passage \cite{stirap}, have been thoroughly investigated with atoms, and recently, also with solid-state quantum dots \cite{EITqd07,EITqd08}. It is intriguing to demonstrate some of these phenomena using macroscopic quantum states in a superconducting-based framework. Phase qubits in particular are well-suited for this purpose because they have a simple ladder-type energy level structure as well as a measurement process that is fast and state specific.

In a phase qubit \cite{martinisqb,cooper04} (see Fig.~1(a)), a single Josephson junction, capacitively shunted with $C_J$, having an effective linear inductance $L_J = \left(\Phi_0/2\pi \right)^2/E_J$, where $E_J$ is the Josephson energy, and $\Phi_0$ is the flux quantum, has been inserted into a superconducting loop. The loop inductance $L > L_J$ is chosen so that local minima are formed in a one-dimensional energy potential $E_{\m{pot}} = (\Phi - \Phi_{\m{ext}})^2/2 L -E_J \cos (2 \pi \Phi/\Phi_0)$ controllable by an externally applied flux $\Phi_{\m{ext}}$. For this experiment, there are approximately ten energy levels residing within a local minimum as shown in Fig.~1(b).
\begin{figure}[h]
\center
\includegraphics[width=8cm]{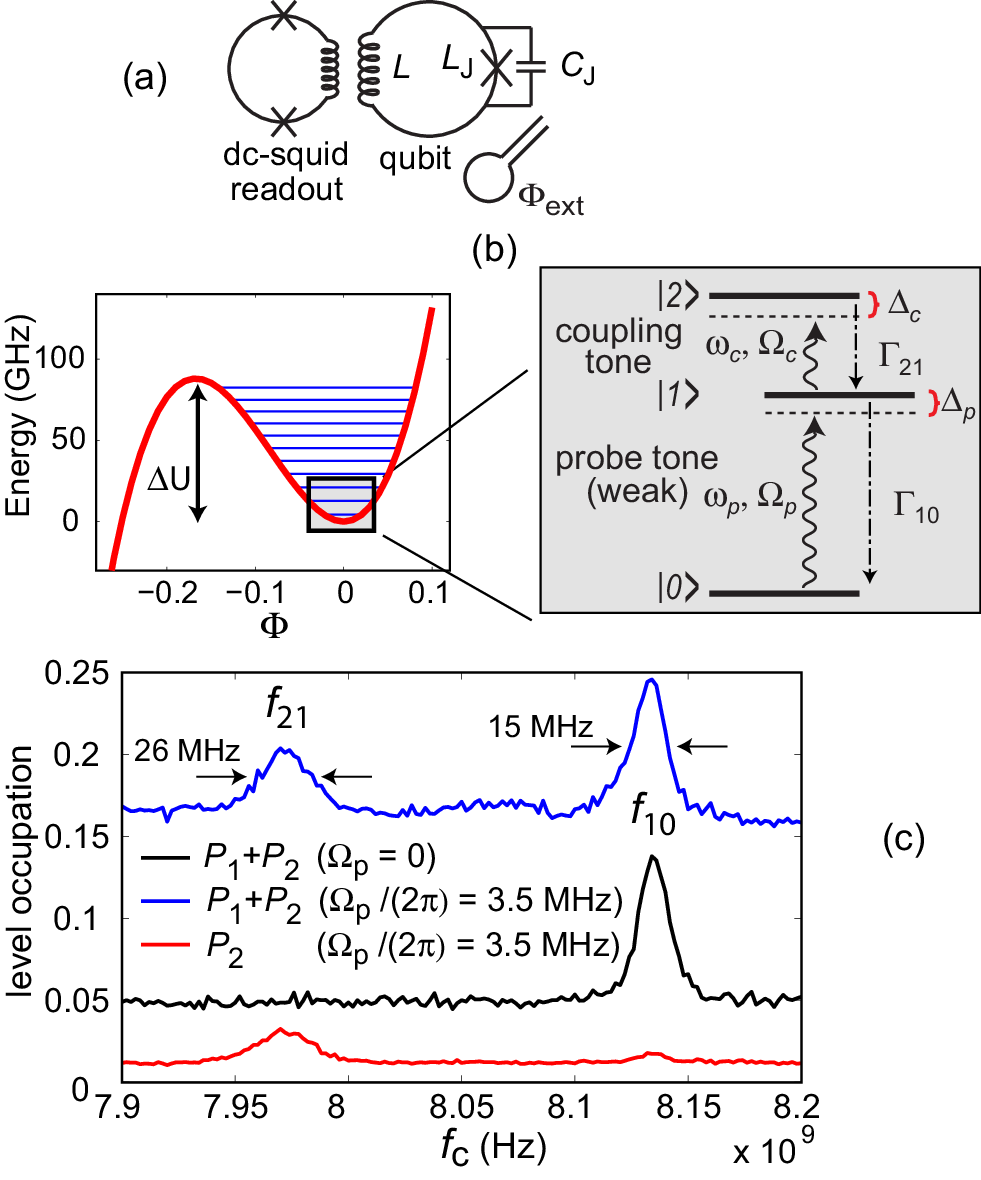}
\caption{(a) Schematics of the phase qubit and a nearby dc-squid for single-shot readout \cite{martinisqb,cooper04} and on-chip flux coil; (b) Energy levels in the local potential minimum at $\Phi \sim \Phi_0$ given for the flux bias point $\Phi_{\m{ext}} \sim -0.42 \, \Phi_0$ used in the experiment. The zoom illustrates the cascade configuration
of the three lowest levels used to generate the Autler-Townes effect. The (strong) coupling field $\omega_c$ connects energy levels $|2\rangle$ and $|1\rangle$, and the weaker probe field $\omega_p$, whose resonant absorption becomes blocked due to the coupling field, connects $|1\rangle$ and $|0\rangle$. Their amplitudes are denoted by $\Omega_c$ and $\Omega_p$, respectively. Generally, both tones are detuned from the respective transitions by $\Delta_c$ and $\Delta_p$; (c) Two-tone microwave spectroscopy to reveal the two lowest transition frequencies $f_{10} = 8.135$ GHz and $f_{21} = 7.975$ GHz (see text for details).} \label{fig1}
\end{figure}

Consider the three lowest energy levels in a ladder or cascade-type configuration in Fig.~1(b). This can be contrasted with the so-called "lambda" configuration familiar in atomic systems, where the state $|2\rangle$ would be lower in energy than the state $|1\rangle$. We denote the transition frequencies between energy levels $i$ and $j$ as $\omega_{ji} = 2 \pi f_{ji}$. Two microwave fields are present; a weak probe tone $\omega_p$ that is nearly resonant with $\omega_{10}$ and detuned by an amount $\Delta_p = \omega_{10} -  \omega_{p}$, and a strong coupling tone $\omega_c$ that is nearly resonant with $\omega_{21}$ (detuned by $\Delta_c = \omega_{21} - \omega_{c}$). The amplitudes $\Omega_c$ and $\Omega_p$ of these tones lead to corresponding Rabi "flopping" angular frequencies denoted by $\Omega_{21}$ and $\Omega_{10}$. In atomic physics, the occupation of higher energy levels will often decay into a continuum of states, but here, they will decay merely to the lower levels. We denote the inter-level relaxation rates by $\Gamma_{21}$ and $\Gamma_{10}$. The transition $|0\rangle \rightarrow|2\rangle$ (rate $\Gamma_{20}$) is a forbidden dipole transition due to the closely harmonic system. We estimate $\Gamma_{20} / \Gamma_{10} \sim$ 6 \%.

In the qubit's eigenbasis $| n \rangle$, the Hamiltonian takes the form $H = \sum_{n=0}^2 E_n | n \rangle \langle n| + \Omega_p \cos(\omega_p t) | 1 \rangle \langle 0| + \Omega_c \cos(\omega_c t) | 2 \rangle \langle 1| +$ h.c. The drive terms arise from inductively coupled microwave flux. In the rotating frame, neglecting counter-rotating terms, the Hamiltonian takes the form
\begin{equation}
H = \left[
  \begin{array}{ccc}
    0 & \Omega_p/2  &   0 \\
    \Omega_p/2      &   \Delta_p & \Omega_c/2 \\
    0 & \Omega_c/2  &   \Delta_p + \Delta_c \\
  \end{array}
\right]\, .
\label{eq:hamilt}
\end{equation}

This Hamiltonian is familiar in atomic physics and has been used to describe coherent population trapping \cite{CohTrapp} and EIT \cite{EIT91}. When both the coupling and probe tones are resonant with their corresponding transitions ($\Delta_c = \Delta_p = 0$), one of the eigenstates of Eq.~(\ref{eq:hamilt}) is the "dark state" $|\Psi_0\rangle = \cos(\Theta) |0\rangle - \sin(\Theta) |2\rangle$, where $\tan(\Theta) = \Omega_p/\Omega_c$, and the system can remain trapped in this state without  $|1\rangle$ ever being populated. Analogous phenomena have been predicted for superconducting qubits \cite{OliverEIT,siewert}, however, experiments have not been forthcoming until now \cite{Wallraff}.

The qubit sample, comprising Al-AlO$_{\m{x}}$-Al Josephson junctions, was fabricated by optical lithography and standard ion-milling techniques on a sapphire substrate. As compared to Ref.~\cite{bus}, the junction size was reduced by a factor of two in order to reduce the number of microscopic TLS defects within the tunnel barrier \cite{simmonds04}. The experiments were performed at 25 mK in a dilution cryostat. The qubit state is measured with a nanosecond-wide flux pulse as described in Ref.~\cite{bus}. The measure pulse reduces the potential barrier $\Delta U$ (Fig.~1(b)) so that the resultant energy level at the top of this barrier, if occupied, has a high probability of tunneling out of the local minimum. The measure pulse amplitude is calibrated properly for investigating excited states $|1\rangle$ and above, when there is a 5 \% residual tunneling probability $P_0$ out of the ground state \cite{cooper04}. This yields a measurement of $P_1 + P_2$, when only the first two excited states are occupied. For investigating excited states $|2\rangle$ and above, a second calibration is performed, by fully populating only the $|1\rangle$ state and establishing a 5 \% residual tunneling probability $P_1$. In our experiments, this provides an independent measure of $P_2$, since all higher states are vacant. Subtracting the results of measurements obtained with both calibrations will always reveal $P_1$ independently.

\begin{figure}[h]
\center
\includegraphics[width=8.0cm]{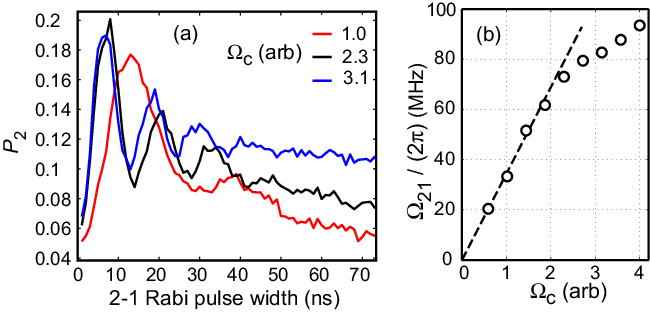}
\caption{Driven Rabi oscillations between energy levels $|2\rangle$ and $|1\rangle$. (a) Excited state probability $P_2$ as a function of the width of a pulse resonant with $f_{21}$ (this pulse follows a $\pi$ pulse resonant with $f_{10}$, see text); (b) the frequency of the Rabi oscillations $\Omega_{21}$ as a function of the coupling drive amplitude $\Omega_{c}$. The dashed line is a fit to the linear portion of the data.} \label{fig2}
\end{figure}
We operate the phase qubit at a center frequency $f_{10} = 8.135$ GHz (Fig.~1(c), black curve) where there are no visible TLS defects over a 1 GHz bandwidth. At high microwave powers, we observe a two-photon absorption peak, $f_{20}/2 = 8.06$ GHz. In order to identify the frequency $f_{21}$, we simultaneously apply two weak microwave tones. We set the probe tone $f_{p}$ to resonantly pump the $|0\rangle \rightarrow|1\rangle$ transition while we vary the coupling tone $f_c$. We obtain a peak in the spectrum when $f_c = f_{21} = 7.975$ GHz as shown in Fig.~1(c). The full spectroscopic data (not shown) as a function of the dc flux bias can be fit well to theory in order to extract the qubit parameters $L_J$ and $C_J$, including any flux offsets.

We also perform time-domain measurements (Fig.~2) by first applying a $\pi$ pulse at $f_{10}$ in order to populate $|1\rangle$ followed immediately by a pulse resonant with $f_{21}$ of varying length. We obtain coherent Rabi oscillations of the second excited state $P_2$, where the population flops between $|2\rangle$ and $|1\rangle$. In Fig.~2(b), we show the Rabi frequency $\Omega_{21}$ of these oscillations for several different coupling drive amplitudes $\Omega_{c}$. According to the ideal three-level model, Eq.~(1), $\Omega_{21} = \Omega_{c}$. A linear relationship between the Rabi frequency and drive amplitude survives up to about $\Omega_{21}/(2\pi) \sim 70$ MHz, indicating that a three-level model is sufficient to describe the dynamics up to this drive \cite{Buisson,Wellstood}.

We made independent measurements of the inter-level decay rates $\Gamma_{21}$ and $\Gamma_{10}$ (Fig.~1(b)) by
preparing populations in either $|2\rangle$ or $|1\rangle$, using sufficiently slow ($\sim 10$ ns wide) $\pi$ pulses
in order to avoid population leakage to the higher levels, and observing energy decay in time domain.
This gave $\Gamma_{21} = 2\pi \times 11$ MHz and $\Gamma_{10} = 2\pi \times 7$ MHz.
The pure dephasing rates can be determined by fitting the spectroscopy peaks in Fig.~\ref{fig1} (c) with simulations of a full model \cite{tbb} with five levels and cross-coupling included;
we obtain $\Gamma^{\varphi}_{10} = 2\pi \times 7$ MHz, $\Gamma^{\varphi}_{21} = 2\pi \times 16$ MHz.
\begin{figure}[h]
\center
\includegraphics[width=8.0cm]{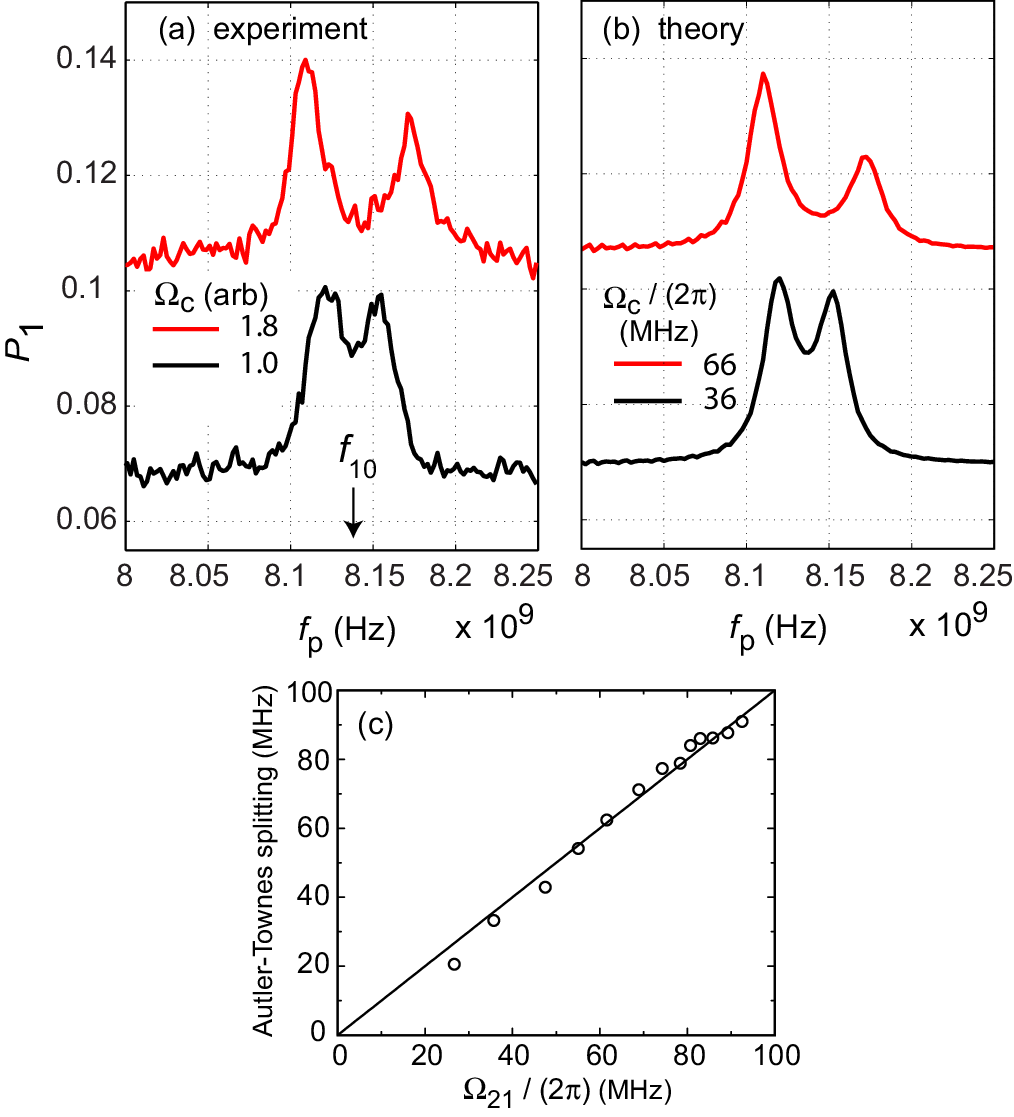}
\caption{Demonstration of the Autler-Townes effect in a multilevel phase qubit system. Plotted is the population $P_1$ of the first excited level at different amplitudes of the coupling field, which cancels the resonant
absorption of a weak probe field $\Omega_p/(2\pi) = 3$ MHz. (a) experiment; (b) theory (full simulation, five levels including cross-coupling); (c) splitting as a function of the independently measured Rabi frequency $\Omega_{21}$ (from Fig.~2(b)): the solid line demonstrates the expected identity between the two quantities as it results from the energy-levels analysis.} \label{fig3}
\end{figure}

As discussed earlier, the "dark state" is expected to occur when both the tones are resonant ($\Delta_c = \Delta_p = 0$). We expect this condition to be visible as a reduced population $P_1$. When $\Omega_c \gg \Omega_p$ the dark state approximates the ground state. We find that the probe tone absorption peak begins to split, forming a dip at $f_p = f_{10}$, see Fig.~3(a). In our cascade level configuration \cite{EITcascade}, energy decay from $|2\rangle$ reduces the effectiveness of the trapping process by populating $|1\rangle$. This hinders our ability to directly produce a completely dark state as a clear reduction in the population $P_1$ or $P_2$. Also, due to a small anharmonicity, the coupling tone drives the $|0\rangle \rightarrow |1\rangle$ transition (see discussion below).

\begin{figure}[h]
\center
\includegraphics[width=8.0cm]{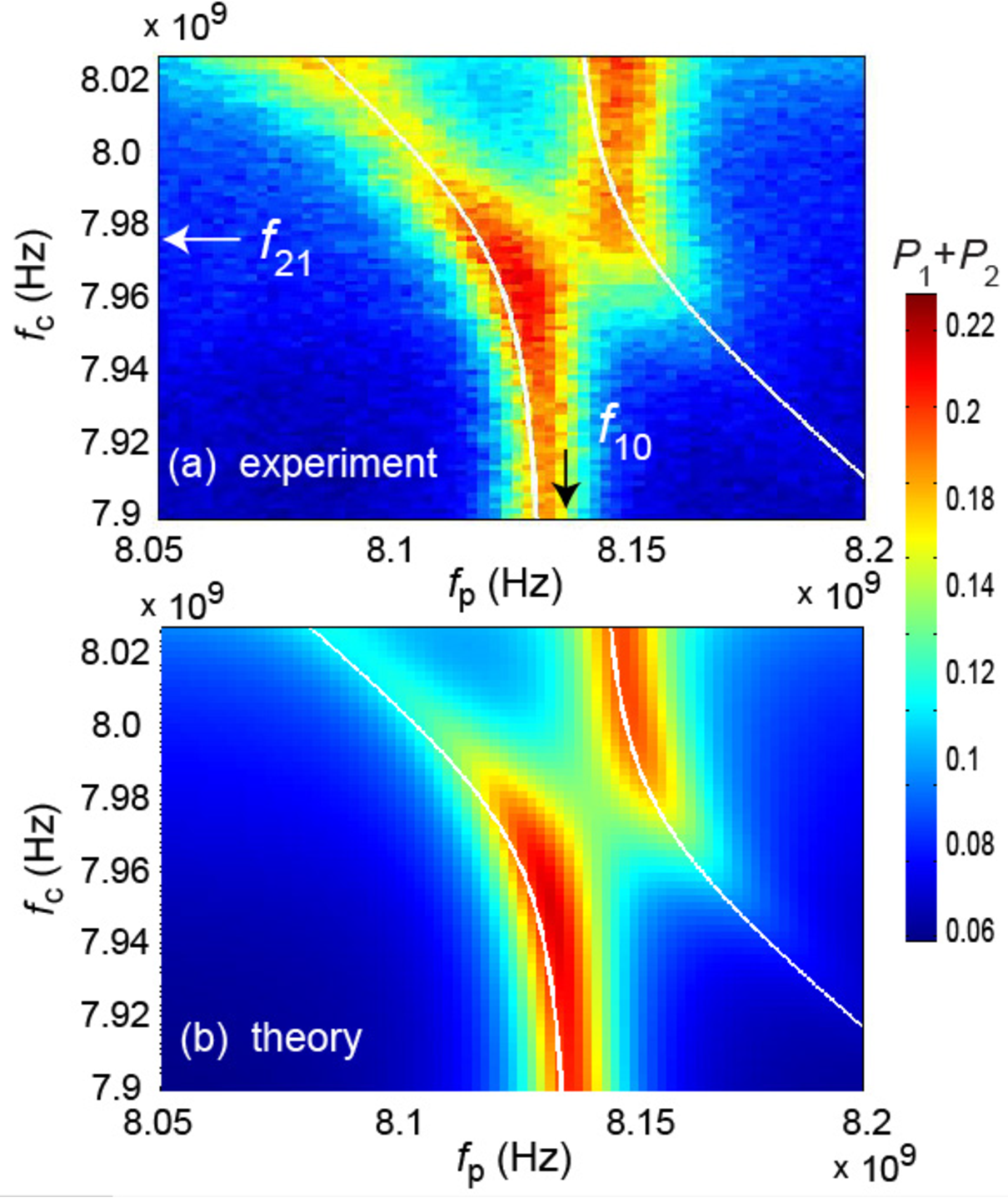}
\caption{Autler-Townes splitting in the three-level phase qubit, for a corresponding coupling field amplitude of $\Omega_c = 1.0$ (in arb units), corresponding to (Fig. 2) $\Omega_{21} /(2\pi) = 36$ MHz, and probe amplitude $\Omega_p/(2\pi) = 3$ MHz. Resonances in the driven system exhibit an avoided crossing as a function of the detuning of the coupling field (vertical axis) and the probe field
(horizontal axis). The white lines mark the expected transitions from the three-level Autler-Townes model. (a) measured excited-level population $P_1 + P_2$; (b) five-level simulation.} \label{figAT33MHz}
\end{figure}

The size of the splitting can be understood in terms of the Autler-Townes effect \cite{AT,ATexp,ATcascade} known in atomic physics. If we treat the amplitude of the probe tone as a small perturbation in Eq.~(\ref{eq:hamilt}), we obtain eigenenergies of the driven system: $\epsilon_{\pm} = \omega_{10} + \Delta_c/2 \pm \sqrt{\Delta_c^2 + \Omega_c^2}/2$. These energy levels are excited from the ground state using the probe tone. When $\Delta_c = 0$, the doublet is spaced by the coupling Rabi amplitude $\Omega_c$. In Fig.~3(c), this splitting is plotted as a function of the independently measured $\Omega_{21}$ (from Fig.~2 (b)), confirming their identity within a 10 \% margin. By scanning the frequencies of the probe and coupling tones, one expects to find that the Autler-Townes doublet will exhibit an avoided crossing centered at $(\omega_{10}, \, \omega_{21})$. We display the results from this type of a measurement in Fig.~\ref{figAT33MHz} with the Autler-Townes eigenvalues plotted as white solid lines. Again we find good agreement for the simple three-level model when operating in the linear regime of the driven Rabi oscillations at $\Omega_{21}/(2\pi) = 36$ MHz (Fig.~2(b)).

The vanishing intensity of the lower right leg of the Autler-Townes avoided crossing in Fig.~\ref{figAT33MHz} can be explained by considering the matrix element $\sin(\phi/2)$ connecting the driven eigenstates with the ground state, where the mixing angle is given by $\tan (\phi) = \Omega_c/ 2 \Delta_c$. As the detuning $\Delta_c$ increases, this matrix element decreases so that the intensity of the split peak on the right also decreases, in agreement with the experimental data. However, the opposite, upper left leg shows a decreasing then increasing intensity with detuning. This can be attributed to an additional two-photon process, which becomes favorable when the probe tone reaches the two-photon resonance at 8.06 GHz, also causing an enhancement of the base level.

For the Hamiltonian of Eq.~(\ref{eq:hamilt}), we have assumed that both microwave fields couple only to their intended transitions, namely the coupling tone $\omega_c$ drives $\omega_{21}$ and the probe $\omega_p$ drives $\omega_{10}$. In reality, the microwave fields cross-couple to both transitions. The strong coupling tone drives the $|0\rangle \rightarrow |1\rangle$ transition, which contributes to the occupation of the first excited state, $P_1 = (1/2)\Omega_c^2 / \left[ \left( \Delta_c + \omega_{10} - \omega_{21}\right)^2 + \Omega_c^2 \right]$, raising the base-level in Fig.~3(b) when far from $f_{10}$. This is contained in a cross-coupling strength $\Omega_c^x$, which for our weakly anharmonic potential, $\Omega_c^x \simeq \Omega_c / \sqrt{2}$. When $\Omega_{21}/(2\pi) = 36$ MHz, (black curve in Fig.~3, Fig.~4), the cross-coupling is negligible. However, for strong coupling amplitudes corresponding to $\Omega_c/(2\pi) = 66$ MHz (red curves in Fig.~3) approaching the limitations of the three-level model, we have to include the cross-coupling.

 So far, our simple three-level model has provided a description of  the Autler-Townes effect in good agreement with the experimental data. Next, we provide a more accurate description based on a full simulation that can account for the multilevel nature of the phase qubit. We numerically solve the standard master equation used to describe such systems \cite{review}, including also in the Hamiltonian cross-couplings of the drive tones up to the fifth level \cite{tbb}. The results are shown along side the experimental data in Fig.~3(b) and Fig.~4(b). The simulation captures the asymmetry of the splitting (Fig.~3(a)), which is a combination of the effects of the higher levels and cross-coupling, as well as the intensity vanishing of the left and right branches as a function of detuning as seen in Fig.~4(b).

In summary, we have observed the Autler-Townes effect, characteristic of three-level systems familiar in atomic physics, but in a superconducting Josephson junction-based quantum system. The results contribute to a general scientific effort that seeks to demonstrate quantum mechanical behavior in progressively more macroscopic and diverse systems. They also pave the way towards quantum information processing using higher-dimensional Hilbert spaces \cite{qudit}.

\begin{acknowledgments}
This work was financially supported by the Academy of Finland. NIST collaborators are supported by NIST and IARPA.
\end{acknowledgments}

\end{document}